\documentclass[preprint,12pt]{elsarticle}

\usepackage{textcomp,booktabs}
\usepackage[usenames,dvipsnames]{color}
\usepackage{colortbl}
\usepackage{rotating}
\definecolor{mygray}{gray}{.9}
\definecolor{mypink}{rgb}{.99,.91,.95}
\definecolor{mycyan}{cmyk}{.3,0,0,0}
\biboptions{numbers,sort&compress}
\usepackage{amssymb}
\usepackage{graphicx}
\usepackage{subfigure}
\usepackage{url}
\usepackage{booktabs}
\usepackage{tabularx}
\usepackage{array}
\usepackage{lscape}
\usepackage{longtable}
\usepackage{multirow}
\usepackage{amsmath}
\usepackage{colortbl}
\usepackage{epsfig}
\usepackage{epsf}
\usepackage{amsthm}

\usepackage[table,xcdraw]{xcolor}
\newtheorem{definition}{Definition}[section]

\usepackage{lineno}
\usepackage{subfig}

\newcommand{\PreserveBackslash}[1]{\let\temp=\\#1\let\\=\temp}
\newcolumntype{C}[1]{>{\PreserveBackslash\centering}p{#1}}
\newcolumntype{R}[1]{>{\PreserveBackslash\raggedleft}p{#1}}
\newcolumntype{L}[1]{>{\PreserveBackslash\raggedright}p{#1}}
\journal{Reliability Engineering \& System Safety} 
\begin{document}

\begin{frontmatter}


\title{The vulnerability of communities in complex network: An entropy approach}



\author[address1]{Tao Wen}
\author[address1]{Yong Deng \corref{label1}}

\address[address1]{Institute of Fundamental and Frontier Science, University of Electronic Science and Technology of China, Chengdu, 610054, China}
\cortext[label1]{Corresponding author at: Institute of Fundamental and Frontier Science, University of Electronic Science and Technology of China, Chengdu, 610054, China. E-mail: dengentropy@uestc.edu.cn, prof.deng@hotmail.com.(Yong Deng)}
\begin{abstract}
Measuring the vulnerability of communities in complex network has become an important topic in the research of complex system. Numerous existing vulnerability measures have been proposed to solve such problems, however, most of these methods have their own shortcomings and limitations. Therefore, a new entropy-based approach is proposed in this paper to address such problems. This measure combines the internal factors and external factors for each communities which can give the quantitative description of vulnerability of community. The internal factors contain the complexity degree of community and the number of edges inside the community, and the external factors contain the similarity degree between chosen community and other communities and the number of nodes outside the community. Considering community vulnerability from the perspective of entropy provides a new solution to such problem. Due to sufficient consideration of community information, more reasonable vulnerability result can be obtained. In order to show the performance and effectiveness of this proposed method, one example network and three real-world complex network is used to compare with some exiting methods, and the sensitivity of weight factors is analysed by Sobol' indices. The experiment results demonstrate the reasonableness and superiority of this proposed method.
\end{abstract}


\end{frontmatter}

\section{Introduction}

Recently, cyber-physical system (CPS) has attracted much attention in numerous fields, such as microgrid \cite{Guan2010Microgrid}, smart city \cite{BRUNEO201912service}, internet of things \cite{Dautov2019Hierarchical,Dautov2018Metropolitan}, and so on. Meanwhile, how to model CPS into specific physical models to analyze their performance and property has become the focus of research \cite{Lee2019cascading}. Therefore, complex network has been applied in this field to better understand the performance of CPS \cite{Cheng2019Smart,Guo2019network}, because nodes in the network represent individuals in the system, and edges would show the relationship between these individuals. Most of previous researches focus on the structure and topological property, which can quantify the characteristic and performance of network. Particularly, the community structure has received increasing attention, because it can reveal human dynamics \cite{wang2018exploiting,wang2017onymity}, inference reliable links \cite{Ma2019Reliable}, and identify influential nodes \cite{ZHANG2019249,wentao2019nodes}. The community structure in the network demonstrates a higher density of nodes and edges, which can cause critical influence on the function and structure of subnetwork, improve the reliability of system \cite{DUI2019162System,LEVITIN2019289Optimization}, and counteract the aging effect \cite{LEVITIN2019397Dynamic,LEVITIN201963Joint}.

There are several different problems about the research of community structure, which can be divided into two issues. The first one is about the structure of community, such as dividing network's community structure \cite{Zhu2019Community,Orman2015Interpreting,Rocco2017Effects}, detecting overlapping community \cite{Orman2015Overlapping}, and dynamic changes of community in evolving network \cite{ORMAN2017375}. Another one is the property of community, including measuring the reliance of community \cite{Ramirez2018Quantifying,Zhang2018uncertainty,Cerqueti2019resilience}, reconfiguring network \cite{Zhang2019Reconfiguration}, quantifying the reliability of community\cite{Ramirez2016Robustness,Zhang2017Reliability}, and measuring community vulnerability \cite{wentao2018evaluating}. The vulnerability of community in network gradually aroused researchers' interest recently. For example, Rocco et al. \cite{Rocco2011Vulnerability} defined vulnerability set and value for different community, and proposed relative vulnerability value to compare with remaining communities. Wei et al. \cite{Wei2018Measuring} proposed a measure which consider more information about community itself, and used non-linear weighted function to combine these factors. Aniko et al. \cite{Kovacs2019vulnerability} proposed a topological index (distance-based fragmentation) to quantify the structural vulnerability in plant-visitation network. Alim et al. \cite{Alim2016Structural} assessed  the community vulnerability through social-based forwarding and routing methods in opportunistic networks, which shows significant contribution about some devices on the performance of entire network. Che et al. \cite{Yanbo2019Vulnerability} modified original evolution method, and proposed a nondimensionalized scoring standard to form a complete assessment system to measure the vulnerability of urban power grid. Chen et al. \cite{CHEN2019CORRELATION} explored the relationship between vulnerability of complex network and fractal dimension. These methods have their own limitations, like computational complexity, inaccurate measurements, and not suitable for certain scenarios.

Since entropy is an useful tool to measure the uncertain of information \cite{Dengwei2018,wang2016statistical}, it has been wildly in the network theory, like dimension presentation \cite{wb2019,WU2018HADAMARD,wentao2018information}, evidence theory \cite{jiang2018Correlation,Jiang2019Znetwork}, influential nodes identification \cite{li2018evidential,wang2018amodified}, time series prediction \cite{xu2018visibility,Fan2019timeseries}. In addition, the structure and property of communities can be expressed by probability sets, entropy-based method has gradually been an reasonable and effective method to quantify the property of network \cite{wentao2019similar,wentao2019structure}. Therefore, an entropy approach is applied in this paper to measure the vulnerability of community which can overcome the shortcomings and limitations of previous method.

In this paper, an entropy-based measure is proposed to quantify the vulnerability degree of community structure. This proposed method can combine two parts of information, i.e., internal factors and external factors, which can consider more information about community and give a reasonable vulnerability result of each community. The internal factors contains the number of edges inside the community and the complexity degree of community which is measured by Tsallis structure entropy, and the external factors contains the number of edges outside the community and the similarity degree between chosen community and other communities which is measured by relative entropy. These two entropy can quantify the property of community more reasonable and effective. Finally, the vulnerability and relative vulnerability result can be obtained by this proposed method to quantitatively describe vulnerability of different community. In order to show the performance and effectiveness of this proposed method, one example network and three real-world complex network are applied in this paper. In addition, the sensitive of four weight factors are analysed by Sobol' indices in Manzi network, and the vulnerability order obtained by different method are compared in Italian 380KV power grid network. The experiments results show the superiority and reasonableness of this proposed method, meanwhile, this proposed method can overcome the shortcomings and limitations of previous method,

The organization of the rest of this paper is as follows. Section 2 presents some basic properties about node and detecting community methods. This novel entropy-based method is proposed in Section 3 to measure the vulnerability of community. Meanwhile, numerical experiments are performed to illustrate the reasonableness and effectiveness of the proposed method in Section 4. Conclusion is conducted in Section 5.

\section{Preliminaries}

In this section, some basic concepts about complex network are introduced. In addition, a community detection algorithm and classical community vulnerability measure are described in this section.

\subsection{Node properties in network}

A given complex network can be denoted as $G(N,E)$, where $N = (1,2, \cdots ,n)$ and $E = (1,2, \cdots ,m)$ is the set of nodes and edges respectively, and $n$ and $m$ is the number of nodes and edges in the complex network respectively. $A$ is the adjacency matrix of complex network whose size is $n \times n$, where ${a_{ij}} = 1$ represents there is an edge between node $i$ and node $j$, and ${a_{ij}} = 0$ is the opposite.

\begin{definition}
(Node Degree). The degree of node $i$ in the complex network is denoted as ${d_i}$ and defined as follows,
\begin{equation}\label{E_degree}
{d_i} = \sum\limits_{j = 1}^n {{a_{ij}}}
\end{equation}
where ${a_{ij}}$ is one element of adjacency matrix $A$. The degree distribution of node $i$ is defined as follows,
\begin{equation}\label{E_deg_dis}
{p_i} = \frac{{{d_i}}}{{\sum\limits_{i \in N} {{d_i}} }}
\end{equation}
\end{definition}

\begin{definition}
(Node Betweenness). The betweenness distribution of node $i$ in the complex network is denoted as ${p_i}^\prime $ and defined as follows,
\begin{equation}\label{E_betweenness}
{p_i}^\prime  = \sum\limits_{s,e \ne i} {\frac{{{g_{se}}(i)}}{{{g_{se}}}}}
\end{equation}
where ${{g_{se}}}$ is the total number of shortest paths between node $s$ and node $e$, and ${{g_{se}}(i)}$ is the number of shortest paths between node $s$ and node $e$ which pass through node $i$.
\end{definition}

\subsection{Community detection algorithm}

Lots of measures have been proposed to detect the community structure in complex network. In order to find the community structure of network, Newman's modularity method \cite{Newman2004Fast} is applied in this paper.

\begin{definition}
(Newman's modularity). For a given complex network $G$ with $k$ communities, the modularity is denoted as $Q$ and defined as follows,
\begin{equation}\label{E_Newman}
Q = \sum\limits_{{c_k} = 1}^k {\left( {\frac{{{e_{{c_k}}}}}{m} - {{\left( {\frac{{De{g_{{c_k}}}}}{{2m}}} \right)}^2}} \right)}
\end{equation}
where $k$ is the number of communities, $m$ is the total number of edges in complex network, ${{e_{{c_k}}}}$ is the number of edges in community ${c_k}$, and ${De{g_{{c_k}}}}$ is the total degree of nodes in community ${c_k}$ which is defined as follows,
\begin{equation}\label{E_Community_degree}
De{g_{{c_k}}} = \sum\limits_{i \in {c_k}} {{d_i}}
\end{equation}
where ${c_k}$ is the set of nodes in community, and ${{d_i}}$ is the degree of node $i$.
\end{definition}

The value of $Q$ can measure the difference of different communities which can show the presence of community structure in complex network. Different value of $Q$ represents different situation. $Q = 0$ means all of the nodes in the network are in one single community and there is no community structure in the network. $Q > 0$ represents there are some kinds of community structure, and $Q = 1$ means the community structure is strong in the network. Meanwhile, Newman and Girvan \cite{Newman2004Finding} suggested the value of $Q$ should fall in the range $0.2 \sim 0.7$, and this value of $Q$ would show the existence of community structures.

The main idea of this method is to find the changes in $Q$, and the step to detect community structure is shown as follows,

\textbf{Step 1}: Each node in the complex network is divided in a single community.

\textbf{Step 2}: Every two communities are integrated into one community in turn, and the value of modularity change $\Delta {Q_{ij}}$ can obtained from the community structure.

\textbf{Step 3}: Community $i$ and community $j$ are integrated into one community with the highest $\Delta {Q_{ij}}$.

\textbf{Step 4}: Repeat Step 2 and Step 3 until $\Delta {Q_{ij}} < 0$.

\subsection{Classical community vulnerability measure}

To measure the vulnerability of community, lots of measures have been proposed. One classical measure is introduced in this section.

\begin{definition}
(Community vulnerability measure). The vulnerability of community $x$ is denoted as ${v_x}$ and defined as follows,
\begin{equation}\label{E_vx}
{v_x} = \frac{1}{{\left| {{V_x}} \right|}},\forall {V_x} \ne \emptyset
\end{equation}
where ${V_x}$ is the set of communities which are connected with community $x$, and $\left| {{V_x}} \right|$ is the number of links which are connected with community $x$.

The relative vulnerability of community $x$ is denoted as ${R_x}$ and defined as follows,
\begin{equation}\label{E_Rx}
{R_x} = \frac{{{v_x}}}{v},v = \mathop {\min }\limits_y ({v_y})
\end{equation}
\end{definition}

\section{The proposed method}

\subsection{Basic method}

\begin{figure}
\centering
\includegraphics[width=15cm]{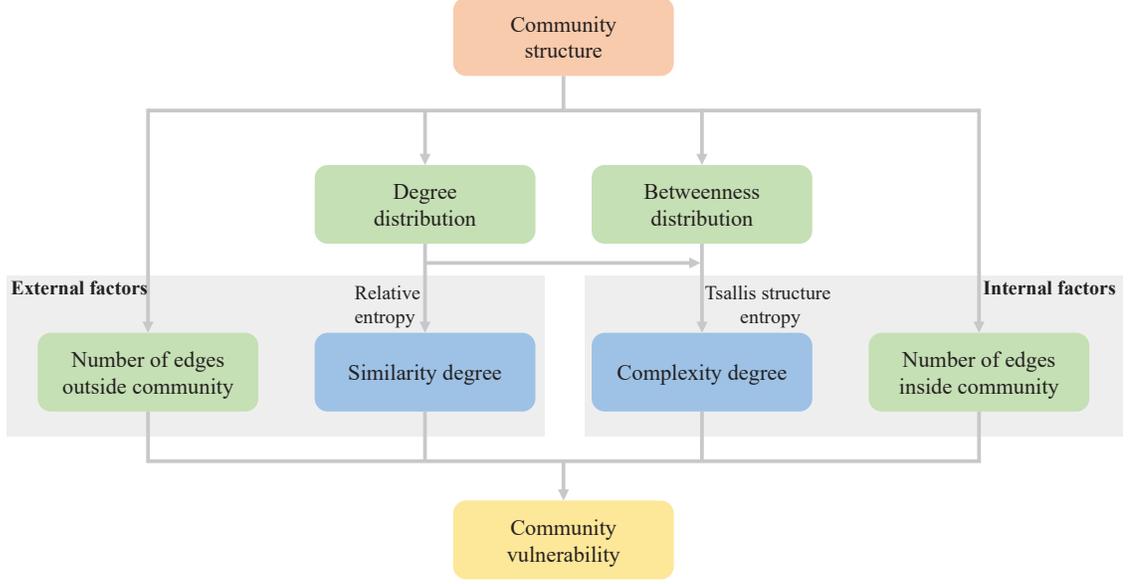}\\
\caption{\textbf{The flow chart of this proposed method.}}
\label{fig_flow chart}
\end{figure}

In this section, a novel method is proposed to measure community vulnerability via entropy approach. This proposed method focuses two parts of information which can consider more details in the network, including internal factors and external factors. The internal factors include the complexity degree of community and number of edges within the community, and external factors include the similarity degree and number of edges between chosen community and other communities. The complexity and similarity would be obtained by entropy method which would overcome the shortcomings and limitations of previous method. The flow chart of this proposed method is shown in Fig. \ref{fig_flow chart}.

\subsubsection{Complexity measure}

Firstly, the complexity degree of community is measured by Tsallis structure entropy which combines the degree distribution and betweenness distribution. Because degree distribution focuses on the local topological information of central node and betweenness distribution considers the global topological information, Tsallis structure entropy which combines these two topological information can give a reasonable measure for community complexity.

\begin{definition}\label{Def_complexity}
(Complexity measure based on Tsallis structure entropy). The complexity degree of community ${c_k}$ is denoted as ${T_{{c_k}}}$ and defined as follows,
\begin{equation}\label{E_complexity}
{T_{{c_k}}} = \sum\limits_{i = 1}^{\left| {{c_k}} \right|} {\frac{{{p_i}^{{q_i}} - {p_i}}}{{1 - {q_i}}}}
\end{equation}
where $\left| {{c_k}} \right|$ is the number of nodes in community ${c_k}$, ${p_i}$ is the degree distribution node $i$ which can be obtained by Eq. (\ref{E_deg_dis}), ${q_i}$ can be obtained from betweenness distribution, and the relationship between ${q_i}$ and ${p_i}^\prime $ is shown as follows,
\begin{equation}\label{E_pq}
{q_i} = 1 + ({p_{\max }}^\prime  - {p_i}^\prime )
\end{equation}
where ${p_i}^\prime $ can be obtained from Eq. (\ref{E_betweenness}), ${p_{\max }}^\prime $ is the maximum value of betweenness ${p_i}^\prime $. The purpose of Eq. (\ref{E_pq}) is to make the index ${q_i}$ bigger than 1 which can show the influence of subnetwork to community ${c_k}$.

When each node's ${q_i}$ equals to 1, Tsallis entropy would degenerate to Shannon entropy which is shown as follows,
\begin{equation}\label{E_Shannon}
{T_{{c_k}}} =  - \sum\limits_{i = 1}^{\left| {{c_k}} \right|} {{p_i}\log {p_i}}
\end{equation}
This form of entropy would only focuses on the local topological structure information.
\end{definition}

The degree distribution is based on the local topological structure around central node $i$. The betweenness distribution focuses on the whole topological structure which can describe the global properties of community. Most of time, ${q_i}$ would be bigger than 1 which can show the influence of subnetwork. When each nodes's ${q_i}$ equals to 1, the Tsallis entropy would degenerate to Shannon entropy based on degree distribution which only focuses on the local structure. Using betweenness distribution to replace the constant parameter $q$ can describe the information about the network itself which is more reasonable for measuring the communities' complexity. This method's property also obeys the classical Tsallis entropy.

\subsubsection{Similarity measure}

Then, relative entropy is used in this section to obtain the similarity degree of chosen community and other communities. The relative entropy (Kullback$ - $Leibler divergence) was wildly used in information theory and probability theory which is proposed by Kullback and Leibler et al. \cite{Kullback1951ON}. In general, the relative entropy is used to measure the difference between two probability set. In this section, the relative entropy is based on Shannon entropy and degree distribution which can measure the similarity between two community structure.

For two community ${c_i}$ and ${c_j}$, the community structure is denoted as ${L_{{c_i}}}({N_i},{D_i})$ and ${L_{{c_j}}}({N_j},{D_j})$ respectively, where ${N_i}$ and ${D_i}$ is the set of nodes and set of degree of nodes in community ${c_i}$. $\left| {{N_i}} \right|$ is the number of nodes in community ${c_i}$ and $\max \left| {{N_i}} \right|$ is the maximum size of community in the network. The probability set of community $i$ is denoted as $P(i)$ and obtained by degree distribution. The scale of every probability set $s$ would be same which equals to $\max \left| {{N_i}} \right|$. So the probability set of community ${c_i}$ can be shown as follows,
\begin{equation}\label{E_probability_set}
P(i) = [p(i,1),p(i,2), \cdots ,p(i,s)]
\end{equation}
The element in probability set is based on degree distribution. When the size $\left| {{N_i}} \right|$ of community ${c_i}$ equals to $\max \left| {{N_i}} \right|$, all of the elements would be obtained by nodes' degree, but when $\left| {{N_i}} \right| < \max \left| {{N_i}} \right|$, some elements would equal to zero to make the probability set complete. The detail of $p(i,k)$ is defined as follows,
\begin{equation}\label{E_element}
p(i,k) = \left\{ {\begin{array}{*{20}{c}}
{\frac{{{d_k}}}{{\sum\limits_{k = 1}^{\left| {{N_i}} \right|} {{d_k}} }}}&{k \le \left| {{N_i}} \right|}\\
0&{k > \left| {{N_i}} \right|}
\end{array}} \right.
\end{equation}
where ${{d_k}}$ is the degree of node $k$, ${\left| {{N_i}} \right|}$ is the number of nods in community ${c_i}$.

To measure the similarity between community ${c_i}$ and ${c_j}$, the relative entropy is used in this section and it is defined as follows,
\begin{definition}\label{Def_similarity}
(Similarity measure based on relative entropy). The difference between two communities is obtained by relative entropy ${R_{ij}}$ and defined as follows,
\begin{equation}\label{E_similarity}
{R_{ij}}(P'(i)||P'(j)) = \sum\limits_{k = 1}^{s'} {p'(i,k)\log \frac{{p'(i,k)}}{{p'(j,k)}}}
\end{equation}
Because the order of element would affect the relative entropy and similarity result, ${p'(i,k)}$ and ${p'(j,k)}$ are the decreasing order of ${p(i,k)}$ and ${p(j,k)}$ in Eq. (\ref{E_probability_set}). ${s'}$ can be obtained as follows,
\begin{equation}\label{E_s_pie}
s' = \min (\left| {{N_i}} \right|,\left| {{N_j}} \right|)
\end{equation}
The adjustment of $s'$ is to avoid ${\frac{{p'(i,k)}}{{p'(j,k)}}}$ being 0 or positive infinity, which would be beneficial for calculation. The relative entropy' property is not symmetry, so the following changes are needed to make it symmetrical,
\begin{equation}\label{E_r_ij}
{r_{ij}} = {R_{ij}}(P'(i)||P'(j)) + {R_{ji}}(P'(j)||P'(i))
\end{equation}
Thus, ${r_{ij}} = {r_{ji}}$ holds, and the relative entropy between two communities are symmetry. Because the relative entropy measure the difference between two probability set, the difference between two communities are obtained in this situation. The bigger ${r_{ij}}$, the greater the difference between two communities structure is. So the similarity index is obtained based on relative entropy to show the similarity between two communities, and it is denoted as ${s_{ij}}$ and defined as follows,
\begin{equation}\label{E_s_ij}
{s_{ij}} = 1 - \frac{{{r_{ij}}}}{{\max ({r_{ij}})}}
\end{equation}
where ${s_{ij}}$ is also symmetry, and shows the similarity between two communities structure. The more similar the two communities, the less the difference between them is, the closer ${r_{ij}}$ is to ${\max ({r_{ij}})}$ and the closer ${s_{ij}}$ is to zero.
\end{definition}

So the similarity between two communities structure can be measured by the relative entropy, which can give a novel approach to this problem. The relative entropy focuses on the local structure topological information in the community structure, which is more reasonable.

\subsubsection{Edges in the network}

The number of edges inside and outside the community are also important for community vulnerability measuring. In this section, the number of edges is considered.

\begin{definition}\label{Def_inside_edge}
(Number of edges inside the community). The number of edges inside the community ${c_k}$ is donated as $D_{{c_k}}^{in}$ and defined as follows,
\begin{equation}\label{E_edges_in}
D_{{c_k}}^{in} = \sum\limits_{i \in {c_k}} {\sum\limits_{j \in {c_k}} {{a_{ij}}} }
\end{equation}
where node $i$ and node $j$ are within the community ${c_k}$, ${a_{ij}}$ is the element of adjacency matrix $A$. Thus, ${a_{ij}}$ is entirely inside the community.
\end{definition}

\begin{definition}\label{Def_outside_edge}
(Number of edges outside the community). The number of edges outside the community ${c_k}$ is donated as $D_{{c_k}}^{out}$ and defined as follows,
\begin{equation}\label{E_edges_out}
D_{{c_k}}^{out} = \sum\limits_{i \in {c_k}} {\sum\limits_{j \notin {c_k}} {{a_{ij}}} }
\end{equation}
where node $i$ is within the community ${c_k}$, and node $j$ is outside the community ${c_k}$, ${a_{ij}}$ is the element of adjacency matrix $A$. Thus, ${a_{ij}}$ connects the chosen community and other communities which can show the relationship between them.
\end{definition}

\subsubsection{Community vulnerability measure}

Lastly, all of the factors defined in Definition \ref{Def_complexity} to \ref{Def_outside_edge} are considered in the vulnerability measuring model. This proposed method would consider the internal factors and external factors which consider more details of community, and is defined as follows,
\begin{definition}
(Proposed community vulnerability measure). The vulnerability of community $x$ is donated as $Vu{l_x}$ and defined as follows,
\begin{equation}\label{E_pro_vul}
Vu{l_x} = \frac{{{{\left( {{S_x}} \right)}^\alpha }}}{{{{\left( {D_x^{out}} \right)}^\beta }}}\frac{1}{{{{\left( {D_x^{in}} \right)}^\lambda }{{\left( {{T_x}} \right)}^\eta }}}
\end{equation}
where $\alpha ,\beta ,\lambda ,\eta $ are the weight factors of different parameters, and all of them are bigger than zero. ${D_x^{in}}$ and ${D_x^{out}}$ is the number of edges inside and outside the community $x$ respectively, ${{T_x}}$ is the complexity degree of community $x$, and ${{S_x}}$ represents the similarity degree between community $x$ and other communities (exclude community $x$ itself) which can be shown as follows,
\begin{equation}\label{E_S_x}
{S_x} = \sum\limits_{j = 1}^k {{s_{xj}}},j \ne x
\end{equation}
where ${{s_{xj}}}$ can be obtained by Eq. (\ref{E_s_ij})

The relative vulnerability of community $x$ is denoted as $R{V_x}$ and defined as follows,
\begin{equation}\label{E_pro_rel_vul}
R{V_x} = \frac{{Vu{l_x}}}{{Vul}},Vul = \mathop {\min }\limits_y (Vu{l_y})
\end{equation}
\end{definition}

In order for these parameters to be considered on the same scale, all of these four parameters ${s_{xj}},D_x^{out},D_x^{in},{T_x}$ are normalized firstly. The weight factors $\alpha ,\beta ,\lambda ,\eta $ can give the weight to consider different parameters, which can be adjusted in different situation. This setting of weight factor makes this proposed method more reasonable. Some special cases of this proposed method $Vu{l_x}$ are shown as follows,

\textbf{1)} When $\alpha {\rm{ = }}\beta {\rm{ = }}\lambda {\rm{ = }}\eta $, this four parameters are considered equally.

\textbf{2)} When $\beta {\rm{ = }}1$, and $\alpha {\rm{ = }}\lambda {\rm{ = }}\eta {\rm{ = }}0$, this proposed method $Vu{l_x}$ would degenerate to the classical vulnerability measure ${v_x}$ in Eq. (\ref{E_vx}).

\textbf{3)} When $\alpha {\rm{ = }}\beta {\rm{ = 0}}$, this proposed method $Vu{l_x}$ would consider the external factors, which is the communities connected with chosen community.

\textbf{4)} When $\lambda {\rm{ = }}\eta {\rm{ = }}0$, this proposed method $Vu{l_x}$ would only consider the internal factors, i.e., the chosen community.

\subsection{Sensitive analysis}

Because these four weight factors ($\alpha ,\beta ,\lambda ,\eta $) are important for community vulnerability measuring, and the vulnerability result would have a related changes as weight factors change, thus, how to determine factors has been a problem in this model. In this section, the sensitive of these weight factors are analysed. In general, the global sensitivity analysis ia a useful tool to obtain the influence of inputs on the output variability in mathematical and physical model, and Sobol' indices based on variance decomposition is applied in this paper. The first-order Sobol' index $SI({X_i})$ and total effect index $ST({X_i})$ are defined as follows respectively,
\begin{equation}\label{E_first_order}
SI({X_i}) = \frac{{Va{r_{{X_i}}}({E_{{X_{ \sim i}}}}(Y\left| {{X_i}} \right.))}}{{Var(Y)}}
\end{equation}
\begin{equation}\label{E_total_effect}
ST({X_i}) = \frac{{{E_{{X_{ \sim i}}}}(Va{r_{{X_i}}}(Y\left| {{X_{ \sim i}}} \right.))}}{{Var(Y)}}
\end{equation}
where $Y$ represents the output of system, ${{X_i}}$ is the \emph{$i$}th independent input $X$, ${{X_{ \sim i}}}$ is all of the inputs exclude ${{X_i}}$, ${Var(Y)}$ is the variance which change with these inputs. The first-order Sobol' index $SI(i)$ can get the contribution of ${{X_i}}$ to $Y$, and total effect index $ST(i)$ can get the contribution to the variance of $Y$ by the variability of each input ${{X_i}}$, which considering its individual effects and the interaction with other variables.

Each weight factors are randomly generated 10000 times by Monte Carlo method, and the range fall into [0.2, 5]. The vulnerability result would be obtained by these random factor combinations, and the contribution of different weight factors can be obtained by first-order Sobol' index and total effect index.

\subsection{An illustrative example}

\begin{figure}
\centering
\includegraphics[width=7cm]{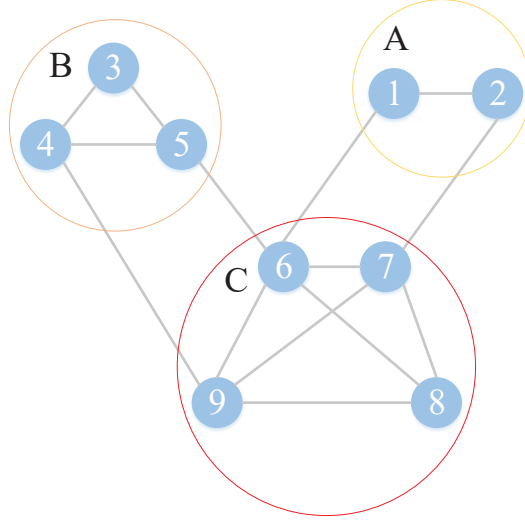}\\
\caption{\textbf{An example network with 9 nodes.}}
\label{fig_example_network}
\end{figure}

In this section, an example network is given to show the difference between this proposed method $Vu{l_x}$ and classical measure ${v_x}$. The network structure is shown in Fig. \ref{fig_example_network}. Observing from Fig. \ref{fig_example_network}, this network has 9 nodes and 14 edges, and the community structure of network is detected by Newman's modularity in Eq. (\ref{E_Newman}). The network is divided into three communities ($Q = 0.2857$) and each of the community is a fully-connected subnetwork. All of these four weight factors $\alpha ,\beta ,\lambda ,\eta $ equal to one which make four parameters equally important. According to this proposed method in Eq. (\ref{E_pro_vul}), (\ref{E_pro_rel_vul}) and classical measure in Eq. (\ref{E_vx}), (\ref{E_Rx}), four parameters and the vulnerability of three communities are shown in Table \ref{table_example_network}.

\begin{table}[]
\centering
\caption{\textbf{The vulnerability of three communities in example network in Fig. \ref{fig_example_network}.}}
\begin{tabular}{cccccccccc}
\hline
Community & ${{S_x}}$ & ${{T_x}}$ & ${D_x^{in}}$ & ${D_x^{out}}$ & ${\left| {{V_x}} \right|}$ & $Vu{l_x}$ & $R{V_x}$ & ${v_x}$ & ${R_x}$\\
\hline
Community A & 1      & 0.5    & 0.1667 & 0.5 & 0.5 & 24     & 27.6264 & 2 & 2\\
Community B & 0.4298 & 0.7924 & 0.5    & 0.5 & 0.5 & 2.1696 & 2.4975  & 2 & 2\\
Community C & 0.8687 & 1      & 1      & 1   & 1   & 0.8687 & 1       & 1 & 1\\
\hline
\end{tabular}
\label{table_example_network}
\end{table}

From Table \ref{table_example_network} when $\alpha {\rm{ = }}\beta {\rm{ = }}\lambda {\rm{ = }}\eta {\rm{ = 1}}$, it can be found that the vulnerability of community C is the lowest, which is the same as the classical measure, but the vulnerability of community A and B is different with classical measure. It can be found from Table \ref{table_example_network} that the classical measure of the vulnerability ${v_x}$ of community A and B is same and it is 2 because of the same number of edges outside of chosen community. However, this classical method is not reasonable, because the vulnerability of one chosen community is determined not only by external factors but also by internal factors. From Fig. \ref{fig_example_network}, community A is a fully-connected subnetwork with only two nodes, but community B is a fully-connected subnetwork with three nodes. When the network' structure is similar, i.e., fully-connected, the network with more nodes would be more robust, so community B is more robust than community A. The similar vulnerability result can be obtained by this proposed method ($Vu{l_B} = 2.1696 < Vu{l_C} = 24$), which is more reasonable for real-world application. The relative vulnerability $R{V_x}$ can be more obvious to show the vulnerability difference between different community. From the comparison result of example network, this proposed vulnerability measure $Vu{l_x}$ outperforms classical method, and can distinguish the vulnerability level of community that the classical method cannot distinguish.

\section{Experimental study}

In this section, three real-world complex networks are applied to show the performance and effectiveness of this proposed method. These three networks are namely as Manzi network \cite{Manzi2001Fishman}, Karate network \cite{Zachary1977Information}, Italian power network \cite{Crucitti2005Locating} respectively. The topological properties of these three networks are shown in Table \ref{table_network_property}. Observing from Table \ref{table_network_property}, $n$ and $m$ is the number of nodes and edges respectively. $\left\langle k \right\rangle $ and ${k_{\max }}$ is the average and maximum value of degree respectively, and $\left\langle d \right\rangle $ and ${d_{\max }}$ is the average and maximum value of shortest distance respectively in the network.

\begin{table}[]
\centering
\caption{\textbf{The topological properties of real-world complex networks.}}
\begin{tabular}{ccccccc}
\hline
Network & $n$ & $m$ & $\left\langle k \right\rangle $ & ${k_{\max }}$ & $\left\langle d \right\rangle $ & ${d_{\max }}$ \\
\hline
Manzi   & 52  & 76  & 2.8077 & 5  & 5.5000 & 13\\
Karate   & 34 & 78 & 4.5882 & 17 & 2.4082 & 5 \\
Italian  & 127 & 171 & 2.6929 & 7 & 8.5682 & 25 \\
\hline
\end{tabular}
\label{table_network_property}
\end{table}

\subsection{Manzi network}

\begin{figure}
\centering
\includegraphics[width=15cm]{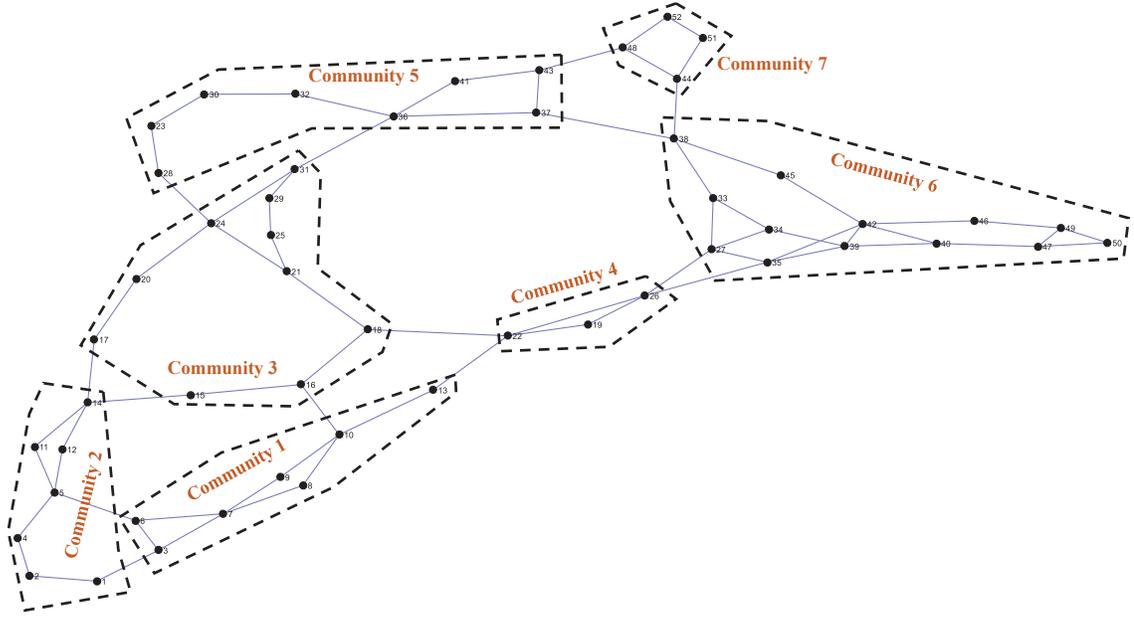}\\
\caption{\textbf{Manzi et al. network \cite{Manzi2001Fishman}.}}
\label{fig_Manzi_network}
\end{figure}

Firstly, the telephone network in Belgium \cite{Manzi2001Fishman} which was analyzed for reliability purposes is used in this section. The topological structure of this network and the community structure obtained by Newman's modularity \cite{Newman2004Fast} is shown in Fig. \ref{fig_Manzi_network}. Observing from Fig. \ref{fig_Manzi_network}, Manzi network is divided into seven communities ($Q = 0.6316$), and the detail nodes in each community is shown in Table \ref{table_Manzi}.

The vulnerability of each community can guide the identification of critical community in the network. The vulnerability $Vu{l_x}$ and relative vulnerability $R{V_x}$ of each community are given in Table \ref{table_vulnerability_Manzi}, the classical comparing vulnerability measure ${v_x}$ and ${R_x}$ are also shown in Table \ref{table_vulnerability_Manzi}. Observing classical measure ${R_x}$ in Table \ref{table_vulnerability_Manzi}, community 7 is the most vulnerable community, but community 3 is the most robust community in the network. The vulnerability of the rest of community (community 1, 2, 4, 5, 6) cannot be identified by the classical measure because of the same value of ${R_x}$. Thus, this novel method is proposed based on two parts of information, including internal factors and external factors. The result of this proposed method ($Vu{l_x}$, $R{V_x}$) are shown in Table \ref{table_vulnerability_Manzi}. It can be found that community 3 is also the most robust community in the network which is same as the classical method. The value of $R{V_x}$ in community 4 and community 7 is close, which are more vulnerable than other communities. The rest of communities can get close but different $R{V_x}$, which can give a vulnerable order of this community (community 6 $>$ community 5 $>$ community 2 $>$ community 1). The vulnerable order of community in the network is community 4 $>$ community 7 $>$ community 6 $>$ community 5 $>$ community 2 $>$ community 1 $>$ community 3. So this proposed method can consider more information in the network and give a detail vulnerable order for these communities which can overcome some limitations of classical method.

\begin{table}[]
\centering
\caption{\textbf{The community details of Manzi network.}}
\begin{tabular}{cl}
\hline
community $k$ & Nodes in community $k$ \\
\hline
1   & 3,6,7,8,9,10,13 \\
2   & 1,2,4,5,11,12,14\\
3   & 15,16,17,18,20,21,24,25,29,31\\
4   & 19,22,26\\
5   & 23,38,30,32,36,37,41,43\\
6   & 27,33,34,35,38,39,40,42,45,46,47,49,50\\
7   & 44,48,51,52\\
\hline
\end{tabular}
\label{table_Manzi}
\end{table}

\begin{table}[]
\centering
\caption{\textbf{The vulnerability of communities in Manzi network.}}
\resizebox{\textwidth}{30mm}{
\begin{tabular}{cccccccccc}
\hline
Community & ${{S_x}}$ & ${{T_x}}$ & ${D_x^{in}}$ & ${D_x^{out}}$ & ${\left| {{V_x}} \right|}$ & $Vu{l_x}$ & $R{V_x}$ & ${v_x}$ & ${R_x}$\\
\hline
1 & 0.3518 & 0.6864 & 0.4210 & 0.6667 & 0.6667 & 1.8260 & 1.8775 & 1.5 & 1.5 \\
2 & 0.3529 & 0.7909 & 0.3684 & 0.6667 & 0.6667 & 1.8170 & 1.8682 & 1.5 & 1.5 \\
3 & 0.4783 & 0.9344 & 0.5263 &   1    &   1    & 0.9725 &   1    &   1 &  1  \\
4 &    1   & 0.5447 & 0.1578 & 0.6667 & 0.6667 & 17.4406 & 17.9322 & 1.5 & 1.5 \\
5 & 0.4139 & 0.8814 & 0.4210 & 0.6667 & 0.6667 & 1.6730 & 1.7202 & 1.5 & 1.5 \\
6 & 0.7654 &    1   &   1    & 0.6667 & 0.6667 & 1.1482 & 1.1805 & 1.5 & 1.5 \\
7 & 0.6429 & 0.5915 & 0.2105 & 0.3333 & 0.3333 & 15.4872 & 15.9238 & 3 & 3\\
\hline
\end{tabular}}
\label{table_vulnerability_Manzi}
\end{table}

Then, Sobol' indices introduced in Section 3.2 is used in this section to analysis the global sensitivity of these four weight factor $\alpha ,\beta ,\lambda ,\eta $. These weight factors can adjust the consideration of different parameters which can give a different vulnerability result. The sensitivity analysis result for the vulnerability of different community with different weight factors are shown in Table \ref{table_sensitivity}. Some conclusions can be obtained as follows,

\textbf{1)} The value of first-order Sobol' index can show the sensitivity of different wight factors. For instance, the vulnerability of community 1 is most sensitive to weight factor $\alpha $, followed by $\lambda$, the other two factors $\beta $ and $\eta$ are less sensitive.

\textbf{2)} When the parameters of community equals to one, the first-order Sobol' index and total effect index would equal to 1. That is because no matter how weight factor change, the influence parameter would remain the same, i.e., equal to 1. For example, ${{T_x}}$ and ${D_x^{in}}$ equal to 1 in community 6, so $SI(\eta )$, $ST(\eta )$, $SI(\lambda )$ and $ST(\lambda )$ equal to 0. Thus, the variability of these two weight factors would not affect the vulnerability measure of community 6. The same situation can occur in ${{S_x}}$ of community 4 and ${D_x^{out}}$ of community 3.

\textbf{3)} In most of communities, the first-order Sobol' index $SI(\beta )$ and $SI(\eta )$ are smaller than $SI(\alpha )$ and $SI(\lambda )$, which means the vulnerability measure of community is more sensitive with $\alpha$ and $\lambda$. The similarity degree and the number of edges within community are more influential to the vulnerability results.

\textbf{4)} The first-order Sobol' index $SI(\beta )$, $SI(\lambda )$, $SI(\eta )$ would be smaller when the value of parameters ${{T_x}}$, ${D_x^{in}}$, and ${D_x^{out}}$ are bigger, and $SI(\alpha )$ is different situation which is bigger with bigger parameters ${{S_x}}$. The situation occurs because $Vu{l_x}$ is positively correlated with ${{S_x}}$, but negatively correlated with ${{T_x}}$, ${D_x^{in}}$, and ${D_x^{out}}$. Hence the value trend of these parameters would have different impact on the vulnerability of communities. These patterns can be observed from all of these communities vulnerability sensitivity analysis in Table \ref{table_sensitivity}.

\textbf{5)} The sum of the first-order Sobol' index over these four weight factors in different communities are less than 1, which means there is an interaction between these four parameters. But this situation does not occur in total effect index.

\textbf{6)} Because of the interaction between these parameters, there would be a huge difference between first-order Sobol' index and total effect index. But it is interesting to find that the order of total effect index would be the same as first-order Sobol' index. For instance, the order of first-order Sobol' index in community 1 is $SI(\eta ) < SI(\beta ) < SI(\lambda ) < SI(\alpha )$, the order of total effect index is $ST(\eta ) < ST(\beta ) < ST(\lambda ) < ST(\alpha )$, which is same as previous order.

\begin{table}[]
\centering
\caption{\textbf{The sensitivity analysis results of the vulnerability of communities in Manzi network with different weight factors.}}
\resizebox{\textwidth}{30mm}{
\begin{tabular}{cccccccccc}
\hline
Community & $SI(\alpha )$ & $ST(\alpha )$ & $SI(\beta )$ & $ST(\beta )$ & $SI(\lambda )$ & $ST(\lambda )$ & $SI(\eta )$ & $ST(\eta )$ & \\
\hline
1 & 0.2055 & 0.7504 & 0.0529 & 0.2604 & 0.1570 & 0.5644 & 0.0294 & 0.2394 \\
2 & 0.1983 & 0.7288 & 0.0509 & 0.2655 & 0.1941 & 0.6308 & 0.0114 & 0.1070 \\
3 & 0.3969 & 0.6861 &    0   &    0   & 0.3116 & 0.5945 & 0.0044 & 0.0124 \\
4 &    0   &   0    & 0.0253 & 0.2524 & 0.4246 & 0.8596 & 0.0883 & 0.4478 \\
5 & 0.2229 & 0.6721 & 0.0650 & 0.2812 & 0.2247 & 0.6074 & 0.0048 & 0.0345 \\
6 & 0.2895 & 0.3756 & 0.6343 & 0.7163 &   0    &   0    &   0    &   0    \\
7 & 0.0275 & 0.3536 & 0.0966 & 0.6666 & 0.1579 & 0.7105 & 0.0374 & 0.3531\\
\hline
\end{tabular}}
\label{table_sensitivity}
\end{table}

Because Sobol' indices is convenient to obtained, and similar results can be obtained from differen network, we only analyse the sensitivity of weight factors in Manzi network and don't analyse the subsequent network.

\subsection{Karate network}

\begin{figure}
\centering
\includegraphics[width=10cm]{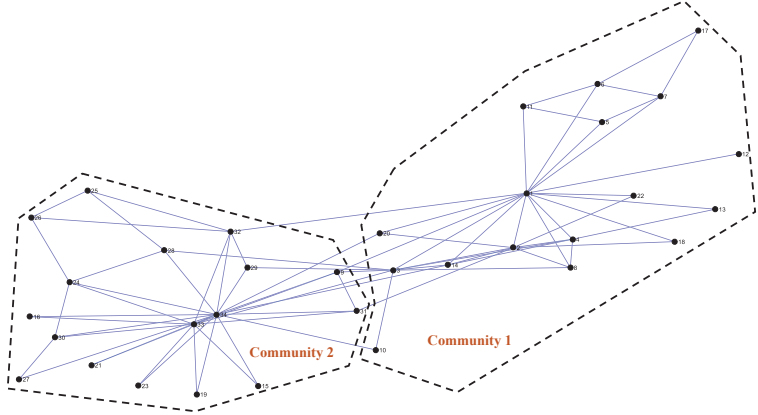}\\
\caption{\textbf{Karate club network \cite{Zachary1977Information}.}}
\label{fig_Karate_network}
\end{figure}

Next, a social network is used in this section to show the performance of this proposed method. This social network is named as Karate club network, which describes the relationship between 34 members of one club in US university \cite{Zachary1977Information}. The topological structure and community structure divided by Newman's modularity is shown in Fig. \ref{fig_Karate_network}. Every nodes in the network denote a member in karate club, including the instructors and administrators, and the edges in the network represent the relationship between two members beyond their normal activities in the club. The network is divided into two communities ($Q = 0.38$), and it is same as the well-known community structure result because there has been disagreement between administrators and instructors \cite{Zachary1977Information}. The detail member in each communities is shown in Table \ref{table_Karate}, and it can be found that the members are divided equally and each community have 17 members. The leader in each community is node 1 and node 34 respectively because of their largest degree.

\begin{table}[]
\centering
\caption{\textbf{The community details of Karate network.}}
\begin{tabular}{cl}
\hline
community $k$ & Nodes in community $k$ \\
\hline
1   & 1, 2, 3, 4, 5, 6, 7, 8, 10, 11, 12, 13, 14, 17, 18, 20, 22 \\
2   & 9, 15, 16, 19, 21, 23, 24, 25, 26, 27, 28, 29, 30, 31, 32, 33, 34 \\
\hline
\end{tabular}
\label{table_Karate}
\end{table}

The vulnerability measure result of Karate network is shown in Table \ref{table_vulnerability_Karate}. Because there is only two communities, it can be found that the external factors (${{S_x}}$, ${D_x^{out}}$) are determined by each other, and they are the same. Thus, the classical measure ${R_x}$ would obtain same result and cannot identify the vulnerability degree of each community. But the internal factors are determined by community itself, these parameters can get a different result. It is interesting to find that the number of edges inside the community ${D_x^{in}}$ is also same and it is 34, but the edges between nodes are different which result in a different complexity degree. The complexity degree ${{T_x}}$ of two communities are 0.7060 and 1 respectively, which would get a different vulnerability measure for different communities. The relative vulnerability $R{V_x}$ of each communities are 1.4162 and 1 respectively, which can get conclusion that community 1 is more vulnerable than community 2. The main reason for their different vulnerability result is the complexity degree of each community, and the initial reason is the topological structure of each community. From this case, we can find that the vulnerability cannot be distinguished when the number of communities is too small. The topological structure inside the community is also important for the vulnerability result, and more factors should be considered to make a accurate identification for their vulnerability. So this proposed method can get a reasonable vulnerability comparison result in Karate club network, whereas, the classical method ${R_x}$ can only get the same vulnerability for two communities.

\begin{table}[]
\centering
\caption{\textbf{The vulnerability of communities in Karate network.}}
\begin{tabular}{cccccccccc}
\hline
Community & ${{S_x}}$ & ${{T_x}}$ & ${D_x^{in}}$ & ${D_x^{out}}$ & ${\left| {{V_x}} \right|}$ & $Vu{l_x}$ & $R{V_x}$ & ${v_x}$ & ${R_x}$\\
\hline
1 & 1 & 0.7060 & 1 & 1 & 1 & 1.4162 & 1.4162 & 1 & 1 \\
2 & 1 &   1    & 1 & 1 & 1 &    1   &   1    & 1 & 1 \\
\hline
\end{tabular}
\label{table_vulnerability_Karate}
\end{table}

\subsection{Italian 380KV power grid}

\begin{figure}
\centering
\includegraphics[width=15cm]{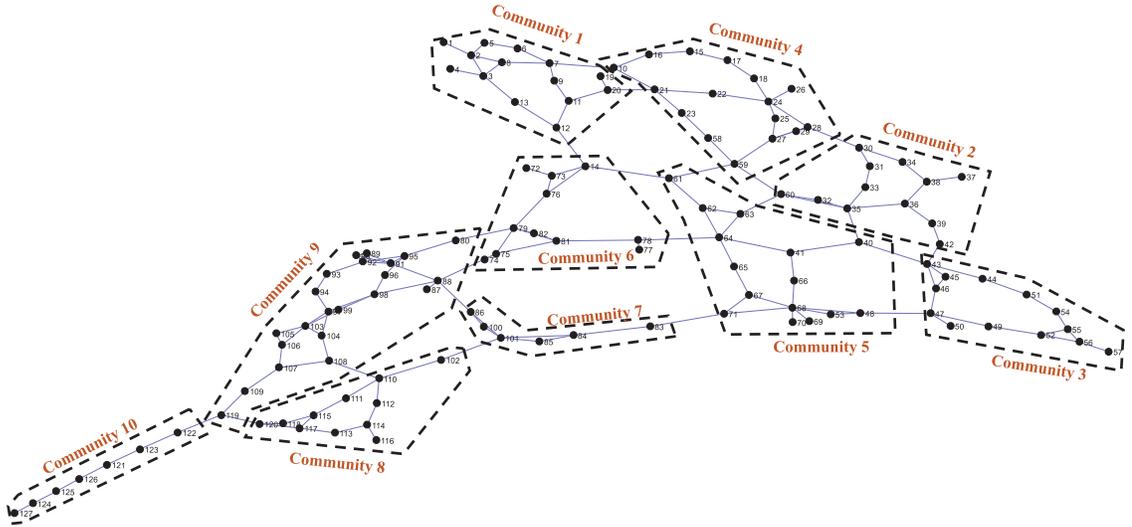}\\
\caption{\textbf{Italian 380KV power grid network \cite{Crucitti2005Locating}.}}
\label{fig_Italian_network}
\end{figure}

Lastly, the Italian 380KV power transmission grid network \cite{Crucitti2005Locating} is used in this section. This network has been frequently used to analyse the network vulnerability performance. The topological structure and community structure is shown in Fig. \ref{fig_Italian_network}. Observing from Fig. \ref{fig_Italian_network}, this network is divided into 10 communities ($Q = 0.7596$), and different communities have different number of components and the detail is shown in Table \ref{table_Italian}. It can be found that community 9 and community 7 have the maximum and minimum number of components respectively.

\begin{table}[]
\centering
\caption{\textbf{The community details of Italian 380KV power network.}}
\resizebox{\textwidth}{30mm}{
\begin{tabular}{cl}
\hline
community $k$ & Nodes in community $k$ \\
\hline
1    & 1, 2, 3, 4, 5, 6, 7, 8, 9, 11, 12, 13, 20, 19\\
2    & 37, 38, 36, 39, 35, 32, 60, 33, 31, 30, 34, 42\\
3    & 57, 56, 52, 49, 50, 47, 46, 45, 43, 44, 51, 54, 55\\
4    & 10, 16, 15, 17, 18, 21, 22, 24, 26, 25, 28, 29, 27, 23, 59, 58\\
5    & 61, 62, 63, 64, 65, 67, 71, 40, 41, 66, 68, 70, 53, 48, 69\\
6    & 77, 78, 81, 74, 75, 79, 82, 76, 72, 14, 73\\
7    & 83, 84, 85, 86, 101, 100\\
8    & 102, 110, 111, 115, 120, 113, 117, 118, 116, 114, 112\\
9    & 119, 109, 107, 108, 106, 104, 103, 105, 97, 99, 98, 88, 87, 96, 91, 95, 80, 92, 90, 93, 94, 89\\
10   & 122, 123, 121, 126, 125, 124, 127\\
\hline
\end{tabular}}
\label{table_Italian}
\end{table}

The vulnerability results are shown in Table \ref{table_vulnerability_Italian}. In this network, a novel method which is proposed by Wei et al. \cite{Wei2018Measuring} is used as a comparing method, this method is also modified from Ref \cite{Rocco2011Vulnerability}. The detail parameters and vulnerability results of communities are shown in Table \ref{table_vulnerability_Italian}. Observing from Table \ref{table_vulnerability_Italian}, community 5 is the most robust community from $R{V_x}$ and ${R_x}$, but ${{R'}_x}$ give a result that community 7 is the most robust. Community 7 only have 6 nodes which is the minimum number of nodes, so this conclusion (community 7 is the most robust) have low credibility, and it can be also seen from these four parameters (${{S_x}}$, ${{T_x}}$, ${D_x^{in}}$, ${D_x^{out}}$) in Table \ref{table_vulnerability_Italian}. Community 10 is consider to be the most vulnerable from these three methods at the same time. This proposed method would magnify the vulnerability of community, like the relative vulnerability of propose method $R{V_x} = 43.8623$ and other method ${R_x} = 8$, ${{R'}_x} = 11.4347$. The $R{V_x}$ of most vulnerable community would be much bigger than other methods, which is convenient to find the vulnerability of community. Meanwhile, the classical method cannot identify some communities' vulnerability, such as community 2, 4, 6 and community 1, 3, 7, 8, because of the same ${R_x}$. This proposed method and Wei et al. method can get a certain vulnerability order for these communities (2, 4, 6 and 1, 3, 7, 8). The detail vulnerability orders of these communities obtained by different methods are shown in Table \ref{table_Italian_order}. From these orders, it can be found that the order obtained by this proposed method is more similar with classical method than Wei method. Specifically, community 9 is considered as the second to last vulnerable community in classical method and proposed method, but it is considered as second vulnerable community by Wei method. Community 3, 7, 8 is considered as the second vulnerable community by classical method at the same time, and this proposed method give a conclusion that community 7, 8, 3 is the second, third, forth vulnerable community respectively which is similar with classical method, but Wei method gives a dissimilar order. Other more detail information about the vulnerability order of communities can be obtained from Table \ref{table_Italian_order}. So this proposed method would consider more information of community and give a certain vulnerability order, and is more reasonable than other methods.

\begin{table}[]
\centering
\caption{\textbf{The vulnerability of communities in Italian 380KV power network.}}
\resizebox{\textwidth}{30mm}{
\begin{tabular}{cccccccccccc}
\hline
Community & ${{S_x}}$ & ${{T_x}}$ & ${D_x^{in}}$ & ${D_x^{out}}$ & ${\left| {{V_x}} \right|}$ & $Vu{l_x}$ & $R{V_x}$ & ${v_x}$ & ${R_x}$ & ${{v'}_x}$ \cite{Wei2018Measuring} & ${{R'}_x}$ \cite{Wei2018Measuring} \\
\hline
1  & 0.3469 & 0.9205 & 0.5161 & 0.3750 & 0.3750 & 1.9470  & 1.9444  & 2.6667 & 2.6667  & 9.1020  & 2.6547 \\
2  & 0.4055 & 0.7882 & 0.4193 & 0.6250 & 0.6250 & 1.9629  & 1.9603  & 1.6000 & 1.6000  & 5.0539  & 1.4741 \\
3  & 0.4675 & 0.8954 & 0.4838 & 0.3750 & 0.3750 & 2.8773  & 2.8734  & 2.6667 & 2.6667  & 6.9333  & 2.0222 \\
4  & 0.5951 & 0.8733 & 0.6129 & 0.6250 & 0.6250 & 1.7789  & 1.7765  & 1.6000 & 1.6000  &  5.6606 & 1.651  \\
5  & 0.4661 & 0.8017 & 0.5806 &   1    &   1    & 1.0013  &    1    &    1   &    1    & 4.0843  & 1.1912 \\
6  & 0.4017 & 0.7588 & 0.4193 & 0.6250 & 0.6250 & 2.0198  & 2.0171  & 1.6000 & 1.6000  & 3.4459  & 1.005  \\
7  &   1    & 0.4891 & 0.2258 & 0.3750 & 0.3750 & 24.1426 & 24.1102 & 2.6667 & 2.6667  & 3.4286  & 1      \\
8  & 0.4114 & 0.8543 & 0.3870 & 0.3750 & 0.3750 & 3.3173  & 3.3128  & 2.6667 & 2.6667  & 8.5554  & 2.4953 \\
9  & 0.7848 &    1   &   1    & 0.7500 & 0.7500 & 1.0465  & 1.0451  & 1.3333 & 1.3333  & 9.5387  & 2.7821 \\
10 & 0.7305 & 0.6875 & 0.1935 & 0.1250 & 0.1250 & 43.9213 & 43.8623 &   8    &   8     & 39.2052 & 11.4347 \\
\hline
\end{tabular}}
\label{table_vulnerability_Italian}
\end{table}

\begin{table}[]
\centering
\caption{\textbf{The vulnerability order of Italian 380KV power network by different method.}}
\begin{tabular}{cl}
\hline
Method & Vulnerability order \\
\hline
Classical \cite{Rocco2011Vulnerability}     & Community 5 $<$ 9 $<$ 2 $=$ 4 $=$ 6 $<$ 1 $=$ 3 $=$ 7 $=$ 8 $<$ 10\\
Proposed              & Community 5 $<$ 9 $<$ 4 $<$ 1 $<$ 2 $<$ 6 $<$ 3 $<$ 8 $<$ 7 $<$ 10\\
Wei et al. \cite{Wei2018Measuring}    & Community 7 $<$ 6 $<$ 5 $<$ 2 $<$ 4 $<$ 3 $<$ 8 $<$ 1 $<$ 9 $<$ 10\\
\hline
\end{tabular}
\label{table_Italian_order}
\end{table}

\section{Conclusion}

The vulnerability measuring of community has already been a hot topic in the study of network theory. In this paper, a new entropy-based method is proposed to measure the vulnerability of communities which can overcome the shortcomings and limitations of previous methods. Different with previous method, this proposed method combines the internal factors and external factors of community which give sufficient consideration of community information. Thus, reasonable vulnerability result can be obtained by this proposed method. The internal factors contain the number of edges inside community and the complexity degree of community measured by Tsallis structure entropy, and the external factors contain the number of edges outside community and the similarity degree between chosen community and other communities measured by relative entropy. The vulnerability and relative vulnerability of community are obtained to give the quantitative description of vulnerability of community eventually. In order to show the performance and effectiveness of this proposed method, one example network and three real-world complex network are applied. Through the vulnerability order obtained by different methods, the rationality of this method is demonstrated. In addition, the sensitivity of weight factors are analysed by Sobol' indices, the important parameters considered in this model can be obtained. The experiment results show the superiority and reasonableness of this propose method.

\section*{Acknowledgment}

The authors thank Prof. Claudio Rocco for providing us with some network data. The work is partially supported by National Natural Science Foundation of China (Grant Nos. 61573290, 61503237).

\section*{Reference}

\bibliographystyle{ieeetr}
\bibliography{myreference}

\end{document}